\documentclass{ws-procs975x65}
\usepackage{graphicx}

\usepackage{etoolbox}

\newtoggle{postrefAA}
\toggletrue{postrefAA}  

\newcommand{\beq}{\begin{equation}}
\newcommand{\eeq}{\end{equation}}
\newcommand{\bea}{\begin{eqnarray}}
\newcommand{\eea}{\end{eqnarray}}
\newcommand{\bit}{\begin{itemize}}
\newcommand{\eit}{\end{itemize}}
\newcommand{\bfi}{\begin{figure}}
\newcommand{\efi}{\end{figure}}
\newcommand{\bfic}{\begin{figure*}}
\newcommand{\efic}{\end{figure*}}
\newcommand{\bce}{\begin{center}}
\newcommand{\ece}{\end{center}}
\newcommand{\bt}{\begin{table}}
\newcommand{\et}{\end{table}}
\newcommand{\btb}{\begin{tabular}}
\newcommand{\etb}{\end{tabular}}

\iftoggle{postrefAA}{
       \usepackage{color}  \definecolor{myred}{rgb}{0.7,0.0,0.2}
}{
      
}

\usepackage{hyperref}

\begin{document}

\title{Inhomogeneous cosmology and backreaction: Current status and future prospects}
\author{Krzysztof Bolejko$^1$ and Miko\l{}aj Korzy\'nski$^2$}

\address{$^1$Sydney Institute for Astronomy, School of Physics, A28,
The University of Sydney, NSW 2006, Australia\\
$^2$Center for Theoretical Physics, Polish Academy of Sciences, Al. Lotnik\'ow 32/46,
02-668 Warsaw, Poland}

\begin{abstract}
Astronomical observations reveal hierarchical structures in the Universe, from galaxies, groups of galaxies, clusters and superclusters, to filaments and voids. On the largest scales it seems that some kind of statistical homogeneity can be observed. As a result, modern cosmological models are based on spatially homogeneous and isotropic solutions of the Einstein equations, and the evolution of the universe is approximated by the Friedmann equations. In parallel to standard homogeneous cosmology, the field of inhomogeneous cosmology and backreaction is being developed. This field investigates whether small scale inhomogeneities via non-linear effects can backreact and alter the properties of the Universe on its largest scales, leading to a non-Friedmannian evolution.
This paper presents the current status of inhomogeneous cosmology and backreaction. It also discusses future prospects of the field of inhomogeneous cosmology, which is based on a survey of 50 academics working in the field of inhomogeneous cosmology.
\end{abstract}



\bodymatter

\section{Introduction}\label{sec:introduction}

In the last decades, we have witnessed significant progress in the field of cosmology. It was fuelled by the increasing amount of observational data on the one hand and
theoretical advancements on the other. The data is now commonly interpreted within the $\Lambda$CDM framework, which assumes that the large--scale geometry of the Universe is described by the spatially homogeneous and isotropic Friedmann--Lema\^itre--Robertson--Walker (FLRW) model with small metric perturbations. These perturbations describe the influence of the matter inhomogeneities and are most commonly treated at the linear level.
The dynamics of the FLRW model is governed by the Friedmann equations with the matter sources assumed to be mostly in the form of dark energy and cold dark matter.
The baryonic component is very small at the level of 5\% of the total energy budget.
This set of assumptions is sufficient to account for most of the cosmological observations, including the anisotropies of the cosmic microwave background, the redshift--luminosity
relation for the type Ia supernovae and various features observed in the distribution of galaxies.
The overall consistency between the observations and predictions of the
$\Lambda$CDM model seems to support the use of the FLRW metric. However, there are a number of reported cases where the $\Lambda$CDM model cannot explain some observational phenomena, or where the constraints on cosmological parameters coming from different experiments seem to be inconsistent with each
other\cite{2016IJMPD..2530007B,2016arXiv161201529C}.

From the point of view of theoretical physics, using spatially homogeneous FLRW models to describe the physical Universe raises some concerns.
Indeed, our Universe on most scales appears to be empty with pockets of matter clustered around various regions and forming the cosmic web.
In order to derive a smooth metric describing the Universe on the large scales we first need to remove the small--scale inhomogeneities. The removal of fine structures and the corresponding degrees of freedom from a complex system is known in theoretical physics as \emph{coarse--graining}. It is a fairly standard procedure in statistical physics or quantum field theory, but in the context of general relativity (GR) seems to be poorly researched.

The main reason for this state of affairs is that the Einstein equations, unlike for example Maxwell's equations, are nonlinear.
In order to obtain a smooth coarse--grained metric
$g_{\mu\nu}^\textit{cg}$, which represents the gravitational field without the contribution from the small-scale inhomogeneities, we first need to average the physical metric $g_{\mu\nu}^\textit{phys}$.
The next step is to average the stress--energy tensor $T_{\mu\nu}^\textit{phys}$ and obtain $T_{\mu\nu}^\textit{cg}$. Assuming that Einstein's equations hold exactly for $g_{\mu\nu}^\textit{phys}$
\bea
G_{\mu\nu}\left[ g_{\mu\nu}^\textit{phys}\right] = 8\pi G T_{\mu\nu}^\textit{phys},
\eea
we can then derive the effective Einstein equations holding on the large scales
\bea
G_{\mu\nu}\left[ g_{\mu\nu}^\textit{cg}\right] = 8\pi G \left(T_{\mu\nu}^\textit{cg} + B_{\mu\nu}\right),
\eea
where $B_{\mu\nu}$ is known as the backreaction \cite{Ellis:2011} and it arises due to the non-linear structure of the Einstein equations.
 In practice we may need more than one application of the coarse--graining procedure to reach the homogeneity scale
 of the solution. The total backreaction in this case will consist of the sum of contributions from all intermediate scales \cite{2005PhLA..347...38E,Ellis:2011, Korzynski:2014nna}.

 The debate on backreaction began in the early 2000's. The discussion still continues and includes such issues as the actual amplitude of backreaction effects and how the backreaction should be estimated. One of the concerns is whether the backreaction requires a fully nonlinear and relativistic treatment or whether the perturbative approach is sufficient\cite{Baumann:2010tm,Ishibashi:2005sj, Green:2010qy, Green:2014aga, Green:2015bma}. The debate seems inconclusive so far; one of the reasons for that is that even if metric perturbations are small their derivatives (and therefore physical curvature\cite{2009GReGr..41.2017B}) can be quite large \cite{Green:2014aga, Buchert:2015iva,  Green:2015bma, Green:2016cwo}.

Other problems are of a more practical and observational nature: how should we pick the background FLRW model? Obviously, it should be
the one which best fits the observational data -- but in what sense exactly? This issue was raised by Ellis in Ref.~\citenum{1984grg..conf..215E} and Ellis and Stoeger in Ref.~\citenum{Ellis:1987zz} as early as the mid 1980s.  It is connected with the question on how the metric inhomogeneities
present on small scales affect the cosmological observations. The precise topics include questions such as:
is the homogeneous background fully justified or does this procedure introduce any bias?
Are the null geodesics in the physical space-time fully mapped into geodesics of the coarse--grained metric?
To what extent should we trust results inferred from observations analysed within a perturbative framework of the FLRW models? And
should we expect to see any deviations from the cosmological predictions of the FLRW models?

All these issues are the subject of research of the inhomogeneous cosmology. In Sec. \ref{sec:status} we provide a short overview of the field. In order to provide a more thorough review of future prospects of this field we decided to approach inhomogeneous cosmologists and ask for their views on the state of research of the inhomogeneous cosmology --- the results of this survey are presented in Sec. \ref{sec:survey}.

\section{Current status}\label{sec:status}

We have decided to divide the field into five categories and discuss the recent
progress in each of them separately.

\subsection{Fundamental problems of coarse-graining}

Before we discuss the observational issues connected with the inhomogeneities we need to address
the fundamental problems of the definition of the background FLRW model and the separation of the
metric and the matter content into the background and the inhomogeneities. This is most easily done by
discussing exact, analytic models in which as many questions as possible may be answered precisely, without the
need for any approximate methods.

The study has been pioneered by Lindquist and Wheeler in 1957\cite{RevModPhys.29.432}, later
also by Wheeler in Ref.~\citenum{springerlink:10.1007/BF01889418}. The authors considered time-symmetric initial data corresponding to a regular lattice of black holes
on a 3-sphere with no other, continuous sources of the gravitational field. The constraint equations in these models can be solved exactly and therefore these models offer the possibility
to investigate the geometry and the properties of universe models with mass concentrated in black holes in the
nonlinear regime, although their time evolution is beyond the analytical methods. 
This approach has recently been revived by Clifton and Ferreira in Ref.~\citenum{2009PhRvD..80j3503C,2011PhRvD..84j9902C,2012PhRvD..85b3502C}. Later the case of all six possible regular configurations of black holes
were discussed thoroughly by Clifton, Rosquist and Tavakol in Ref.~\citenum{Clifton:2012qh}, while
Korzy\'nski in Ref.~\citenum{Korzynski:2013tea} considered arbitrary arrangements of black holes, discussing the backreaction and the continuum limit
as the number of BH's diverges. In the follow-up Clifton in Ref.~\citenum{Clifton:2014mza}  used the method of images to construct similar solutions with
BH's as the only source of the gravitational field and discussed the contribution of the interaction energy between the BH's to the energy budget of the cosmological model.

The first paper discussing the time evolution of the 8BH regular lattice on $S^3$, Ref.~\citenum{Bentivegna:2012ei}, has been written by Bentivegna and Korzy\'nski. Shorly afterwards Yoo {\em et al.}
discussed the regular cubic black hole lattice in Ref.~\citenum{Yoo:2013yea}. The latter case is
more difficult in the sense that the constraint equations
  need to be solved numerically. A method
  for solution was proposed in an earlier paper
by Yoo, Abe and Nakao\cite{Yoo:2012jz}. The results of the simulations are consistent among the papers: the evolution follows
with a reasonable accuracy the evolution of an appropriately fitted FLRW model with dust.
The flat, cubic lattice has been considered again in Ref.~\citenum{Bentivegna:2013jta} by Bentivegna and Korzy\'nski, who investigated in more detail the dynamic and kinematic effects of backreaction, discovering high frequency oscillating modes superimposed on the FLRW behaviour, attributed to the tensorial modes
contained in the initial data. Later Yoo and Okawa in Ref.~\citenum{Yoo:2014boa} considered cubic BHL's  with a cosmological constant, once again noting a reasonable agreement with
the corresponding FLRW model.

Later Clifton, Gregoris, Rosquist and Tavakol in Ref.~\citenum{Clifton:2013jpa} tried a simplified approach -- instead of evolving the whole spacetime they focused on special loci, like the
vertices and edges of the lattice. In these points the geometry exhibits a discrete rotational symmetry, i.e., it is invariant with respect to a finite subgroup of the
full $SO(3)$ group of rotations. The Einstein equations can be simplified in this case: in the vertices they can be integrated exactly, while along the edges, where the symmetry group
is smaller, they can be solved only as long as the magnetic Weyl contribution is neglected\cite{Korzynski:2015isa,Clifton:2016mxx}. Nevertheless the authors managed to obtain a reasonable approximate
picture of the evolution of these models, valid for a reasonably long time, at least along certain distinguished curves in the spacetime.
The authors confirmed that the evolution of these models follows an appropriately fitted FLRW model with a good accuracy, at
least as long as the approximation is valid.

\subsection{Exact inhomogeneous cosmological solutions}

Another approach of studying the effects and impact of inhomogeneities, is to investigate the evolution of exact
inhomogeneous solutions of the Einstein equations.
There is only a handful of such solutions that are applicable to cosmology\cite{1997icm..book.....K,Bolejko:2011jc}.
The most widely used solutions are the spherically symmetric Lema\^itre--Tolman models\cite{Lemaitre,LemaitreReprint,Tolman,TolmanReprint} (also known as Lema\^itre--Tolman--Bondi, or in short LT or LTB models), and the Szekeres models\cite{Szekeres:1975ct, Szafron:1977zza}.
The Lema\^itre--Tolman models are spherically symmetric. On the other hand the matter distribution in the Szekeres model (on a surface of constant $t$ and $r$) has a form of a dipole superimposed on a monopole\cite{2006igrc.book.....P}. The Szekeres solution is a generalization of the  Lema\^itre--Tolman model and in the limit of vanishing dipole, it reduces to the  Lema\^itre--Tolman model.
Although these models are subject to some limitations (the magnetic part of the Weyl tensor is zero, Cotton--York tensor is zero, and they are of Petrov type D) they allow us to study non-linear effects and structure formations. For example due to presence of shear (which is a nonlinear quantity)
the structure formation proceeds more rapidly than in the linear regime.\cite{2007PhRvD..75d3508B} In addition, the Lema\^itre--Tolman and Szekeres models have both the FLRW and the Schwarzschild limits,  and therefore, these models are useful to study the formation of cosmological black holes and their properties
either in the case of a spherical symmetric collapse\cite{2004PhRvD..69d3502K} or anisotropic collapse\cite{2012PhRvD..85l4016K}.

However, in the contexts of backreaction these models have limited applications. The problem is the choice of the background. In a typical application one
chooses a background (for example, some FLRW model) and on the top of this background one imposes an inhomogeneity.
In such a case, the evolution of the model is already fixed. The junction conditions explicitly require that the evolution
of the boundary of the inhomogeneous model must be the same as of the background. Even if one uses asymptotic conditions (i.e., for large $r$ the model reduces to FLRW model), still the overall evolution of such a system is close to the background FLRW evolution (even if locally inhomogeneities evolve in a nonlinear manner).
Although backreaction may accumulate from scale to scale\cite{Korzynski:2014nna}, within these models the backreaction is still small by  construction. This was shown for the LT model\cite{2011JCAP...05..003M,2011CQGra..28p4004M}. For the Szekeres model with the central observer, the averaging wipes out the dipole and the result of averaging leads to the  same conclusions as for the Lema\^itre--Tolman model\cite{2009GReGr..41.1585B}.

In order to obtain large backreaction or large observational effects
for these types of models
one needs to consider ultra large scales, i.e., of order (or even larger) of Gpc scales.
These models have large backreaction and large deviations from the FLRW distance-redshift relation\cite{2000A&A...353...63C,2006PhRvD..74j3520A,2007JCAP...02..019E,2007PhRvD..75b3506A,2008PMCPA...2....1B,2008PhRvL.101y1303Z,2008JCAP...04..003G,2008JCAP...09..016G,2009JCAP...02..020B,2010JCAP...11..030B,2010A&A...518A..21C,2011JCAP...02..013C,2011PhRvD..83f3506N}, however, they are not consistent with observational constraints and therefore cannot be consider as suitable models of our Universe.\cite{2010PhRvD..82j3532F,2011PhRvD..83j3515M,2012PhRvD..85b4002B,2012JCAP...10..009Z,2014A&A...570A..63R} 

The investigation of the impact of small-scale inhomogeneities on light
propagation in exact solutions shows that the impact in these cases
is rather small, of order of a few percent\cite{2012JCAP...05..003B}. Such a few percent change could in principle lead to a few percent shift in
the values of the
cosmological parameters\cite{2011JCAP...02..025B,2011A&A...525A..49B}.
The exact magnitude of deviation and shift of values of cosmological parameters is still debatable. The reason for this is the lack of inhomogeneous models that would faithfully represent our Universe.

However, studies of light propagation within inhomogeneous models do lead to interesting results and new discoveries, such as an effect of a {\em position drift} --- due to inhomogeneous nature of the Universe, photons sent at different times traverse along different paths (with different matter distributions along the line of sights), which results in a stray shift on the position of these object in the sky.\cite{2011PhRvD..83h3503K}

From the point of view of backreaction, exact solutions provide a useful testbed as they can be
used to solve null geodesics,  matter evolution, and evaluate averages in an exact manner\cite{2011CQGra..28w5002S,2013CQGra..30w5001S,2013PhRvD..87l3503B}. For example, studies of the deceleration parameter showed that the value of this parameter inferred from light tracing can differ significantly from the value obtained via averaging, and only in the FLRW limit these two quantities converge\cite{2008JCAP...10..003B,2010GReGr..42.2453K}.
Unfortunately, due to limitations mentioned above, there is a great need for less restrictive and more general models (cf. Refs.~\citenum{2015PhRvD..92h3533S,2016JCAP...03..012S} Sussman {\em et al.}), which would allow for more extensive investigation and deeper understanding of the backreaction phenomenon.

\subsection{Metric of the Cosmos and testing the homogeneity of the Universe}

The idea of deriving the metric of our Universe strictly from observations  was first suggested by Kristian and Sachs in Ref.~\citenum{1966ApJ...143..379K}.
This is in principle a very promising approach. Once fully developed it would allow for constraining the spacetime geometry directly from the data, rather than fitting the assumed metric to the data. This program is still in its infancy but it has delivered a few interesting developments, such as the fluid-ray tetrad formalism\cite{1992CQGra...9..493S}. Most of the work so far has been based on a framework that assumes spherical symmetry\cite{2007CQGra..24.4107H,2008PhRvD..78d4005M,2009PhRvD..79d3501H,2011JCAP...09..011B}. Still, spherically symmetric inhomogeneous models do have a homogeneous limit. Therefore, if the observational data constrain the spacetime to be homogeneous and isotropic that would be an independent check of the homogeneity of the Universe. In principle this provides means of testing the homogeneity of the Universe, rather than assuming it. The works are still being developed with the aim of putting tighter constraints on the spacetime geometry\cite{Bester:2013fya, Bester:2015}.

A less ambitious alternative of the program (but nonetheless also very important) is rather than constraining the metric of our Universe from the data, to use the data to directly test fundamental assumptions of the standard cosmological model, which is based on the spatially homogeneous and isotropic FLRW geometry. One way is to study the distribution of galaxies \cite{2011CQGra..28p4003S,2016arXiv161102139P}. 
The other method often used is based on the evaluation of the fractal dimension
\cite{2009arXiv0910.4877R,2012MNRAS.425..116S,2015PhyA..417..332C}.
This however, might be  subject to some bias either observational\cite{2011CQGra..28p4003S}
or theoretical\cite{2005A&A...429...65R,2008A&A...488...55R}. Therefore a more promising approach seems to focus on observables such as distance-redshift and expansion rate\cite{2008PhRvD..78j3502S,2008PhRvL.101r1301Z,2008PhRvL.101a1301C} or just distance and bi-distance\cite{2015PhRvL.115j1301R}. If the FLRW geometry holds then these observables are related to each other, and so their intrinsic relations can be used to test the FLRW geometry. So far the observational data is consistent with the FLRW predictions but it is expected that just in a few years (with data from the DES and Euclid surveys) the observations should be precise enough to
show deviations from or to
confirm to a higher accuracy the fundamental assumptions of the FLRW models\cite{2009PhRvD..79h3011L,2009PhRvD..80l3512W,PhysRevD.90.023012,2015PhRvL.115j1301R}.

\subsection{Approximate methods}

Most cosmologists agree that the evolution of the Universe around the decoupling instant and shortly afterwards can be described using the linear perturbations around the FLRW background. However, it is also known that quite quickly the evolution of structures becomes non-linear. In order to trace the evolution in the nonlinear regime, cosmologists either employ perturbative methods or use Newtonian $N$-body simulations.

As far as the perturbative approach is concerned, most efforts focus on the non-linear evolution of the matter power spectrum\cite{Baumann:2010tm,2015JCAP...05..007B,2015PhRvD..92l3007B,2015arXiv151205807B}. However, the perturbative approach may not be accurate if the backreaction is large and affects the evolution of the background\cite{2009PhRvD..80h3525C}. The same concerns $N$-body simulations that assume a uniform Friedmannian cosmological expansion. 
Therefore, investigation of relativistic affects within $N$-body simulations is important\cite{2013PhRvD..88j3527A,2014CQGra..31w4006A,2016JCAP...07..053A}. One of the recent studies suggests that relativistic effects may lead to Yukawa-type interactions between matter particles\cite{2016ApJ...825...84E}, which may have repercussions for $N$-body simulations that assume instantaneous Newtonian interactions.

Studies of backreaction within the perturbative schemes are also slowly being developed\cite{2012PhRvD..86b3520B,2013PhRvD..87l3503B,2015PhRvD..92b3512A}. In parallel, there are also phenomenological models inspired by backreaction, which could in principle also be employed within N-body simulations\cite{2016arXiv160708797R}.
However, there is a problem with studies of backreaction based on approximate or phenomenological methods.
Some studies suggest large backreaction\cite{2006JCAP...11..003R,2008JCAP...04..026R,2007NJPh....9..377W,2009PhRvD..79h3011L,2009PhRvD..80l3512W,2010PhRvD..82b3523W,2011CQGra..28p4006W,2013JCAP...10..043R,2016arXiv160806004R}, while some suggest otherwise\cite{Green:2010qy, Green:2014aga, Green:2015bma}.
The debate on whether the backreaction is zero has already been settled\cite{Buchert:2015iva}.
The current debate is on the magnitude of the backreaction, i.e., whether inhomogeneity effects are in the percent range or of order of unity. It seems now very likely that this debate will eventually be solved using numerical relativity rather than phenomenological or perturbative methods.

\subsection{Numerical cosmology}

Since the number of exactly solvable models in which we may study the time evolution of inhomogeneous spacetimes is very limited, it is natural to consider
 numerical methods of evolving the full Einstein equations in order to study the behaviour of strongly inhomogeneous models. Recent advancements in the field of
numerical relativity\cite{Cardoso:2014uka}, including the possibility to evolve vacuum multi-blackhole configurations,
make this approach particularly tempting. The simplest models of this kind which were evolved using numerical methods were the black hole lattices mentioned in the previous section.

The next development in numerical cosmology is to consider inhomogeneous distribution of dust instead of black holes, which is a more realistic situation than a black hole lattice, although the numerics involved is much more difficult. Bentivegna and Bruni in Ref.~\citenum{Bentivegna:2015flc} considered
a perturbation of an Einstein-de Sitter model with the initial density contrast of $\delta \approx 0.03$, which developed quickly well beyond the linear regime,
and they noted departures in the evolution from the simplified top-hat model. Mertens, Giblin and Starkman in Ref.~\citenum{Mertens:2015ttp} on the other hand specified a spectrum of
scalar perturbations and compared the lengths of a number of selected paths with the uniform FLRW expansion.
Most recently, Macpherson, Lasky, and Price, presented their results of numerical cosmology in Ref.~\citenum{2016arXiv161105447M}. Their study was based on the Cactus code and they were able to trace the evolution inhomogeneities into the nonlinear regime.

The field of numerical relativistic cosmology is slowly
gaining momentum. It is hoped that in a few years, numerical cosmology will allow to test various assumptions of the standard cosmology based on the FLRW models and will deepen our understanding of backreaction.

\section{Future prospects --- survey of the community} \label{sec:survey}

The number of people actively working in the field of inhomogeneous cosmology is of the order of a hundred, which is small compared to at least a few thousands working in the paradigm of the FLRW models (the Euclid team alone counts more than a thousand people). In order to get a better understanding of the community we designed a short survey.
We approached approximately 70 academics, from which 50 have responded. This gave us a good representation both in terms of demographics (from very senior and highly regarded members of the community
to junior academics) and geography (every continent except for Antarctica).

The survey consisted of the following 3 questions:

\begin{enumerate}
\item
What do you think is the most important topic (or range of topics) in cosmology in general?
\item
Within the domain of inhomogeneous cosmology which topic (or topics) are in your opinion the most important?
\item
 How do you think the field of inhomogeneous cosmology will (or should) develop over the next 5--10 years?
\end{enumerate}

Below we present a short summary of the answers we received. The responses were grouped into specific bins. While this approach is not perfect and some of the bins overlap, it enabled us to report the responses in a rather compact way. Also, when a particular answer was unique and not shared by any other respondent we decided not to include such a response in the analysis below.

\subsection{ Question 1: What do you think is the most important topic (or range of topics) in cosmology in general?}

We first asked about the most important topic in cosmology in general, see Fig. \ref{q1-fig}. When compared with answers to {\em Question 2} and {\em Question 3} this will show how the inhomogeneous community perceives its place within the broader frame of general cosmology. The answers we received are as follows:

\begin{figure}[h]
\begin{center}
\includegraphics[scale=0.245]{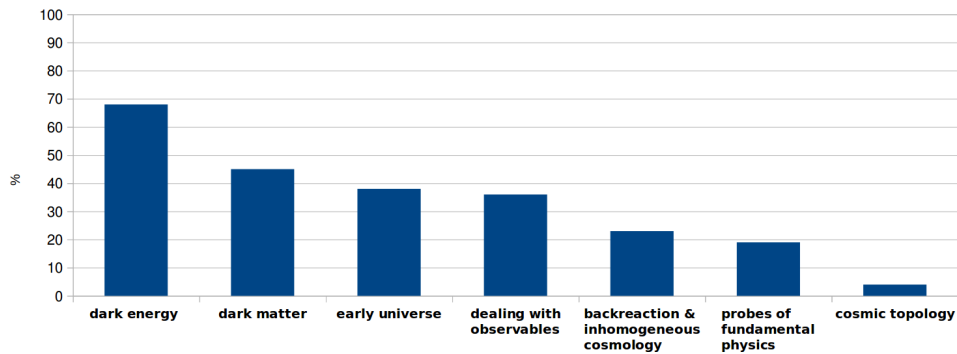}
\caption{{\em What do you think is the most important topic (or range of topics) in cosmology in general?}}
\label{q1-fig}
\end{center}
\end{figure}

\paragraph{Dark sector:}

The most important issue for the majority of respondents was to reveal the nature of dark sector and the fundamental physics behind it. Most pointed out that we need to work in both directions: experimental, with a direct detection of dark matter and dark energy, and theoretical, with better theories exploiting the properties of the dark sector. We also need to examine if these phenomena are related to each other.
While most people answered that both dark energy and dark matter are
equally important, some made a clear distinction,
claiming that since dark energy is most likely the cosmological constant, it is not a
  highly important research goal and that instead, it is dark matter that should take
precedence, as it leads to synergy between cosmology and particle
physics. Others though pointed out that dark energy being a
cosmological constant would require an improbable amount of fine
tuning and is unnatural from the point of view of quantum theory and
thus studies of dark energy could open a new window to fundamental
physics.
Whether dark energy and dark matter are real phenomena or
merely phenomenological descriptions, we still need better models.  Reducing these phenomena to just two numbers ($\Omega_m$ and $\Omega_\Lambda$) is unacceptable.

\paragraph{Early Universe:}

More than a third of the respondents mentioned the early Universe. The majority pointed out weakness of inflation, and that we need other alternative theories describing conditions and problems related to the early Universe. Areas of research that were mentioned included: initial conditions for the Universe, inflation, re-heating, baryogenesis and baryon asymmetry. It was also stressed that the early Universe opens a new window to fundamental physics.

\paragraph{Dealing with observations:}

36\% of respondents pointed out that we had entered an era where a very large number of high precision data became or soon would  become available. New types of measurements are also coming online, for example gravitational waves. Soon we will be able to measure the expansion of the Universe via the redshift drift, and probe the dark ages and the epoch of reionization with 21cm observations. Together with deep and wide galaxy redshift surveys and lensing surveys all this data will undoubtedly lead to new discoveries and perhaps to a paradigm shift.
This will also bring new challenges. It will require a better understanding of various physical processes and knowledge how they impact observables (it has happened a few times in the past that some ``discoveries'' were announced prematurely, and
when a more realistic description of astrophysical processes was included, they went away).
We will also need to learn how to deal with such a large amount of data. The usual approach is to compress the data by projecting it onto a smaller number of functions or parameters. Care needs to be taken in order to perform such a reduction in rather model independent way and without introducing model assumptions into the compressed data and the actual analysis itself. Thus a lot of effort should be put towards scrutinising and falsifying various aspects of the concordance cosmology. With all these new probes and data, care should also be taken to extract information in a model-independent manner and in addition to be aware of the predictions of alternative models. Without
this, any debate on ``precise versus accurate'' cosmology may be inconclusive.

\paragraph{Backreaction and inhomogeneous cosmology:}

Almost a quarter of responses pointed out that the main challenge of modern cosmology is to better understand the effects associated with the inhomogeneous structure of the Universe and backreaction.

\paragraph{Cosmological probes of fundamental physics and modified gravity:}

Almost a fifth of respondents pointed out that cosmology provides a great opportunity to explore new connections between cosmology and fundamental physics. It was noted that cosmological observations can be used to test laws of nature and its fundamental components: from the nature of dark matter, dark energy, and conditions of the early Universe to modified gravity.

\paragraph{Cosmic topology:}

A few people argued that one of the important issues of modern cosmology is to find out what is the overall topology of the Universe we live in.

\subsection{Question 2: Within the domain of inhomogeneous cosmology which topic (or topics) are in your opinion the most important?}

\begin{figure}[h]
\begin{center}
\includegraphics[scale=0.245]{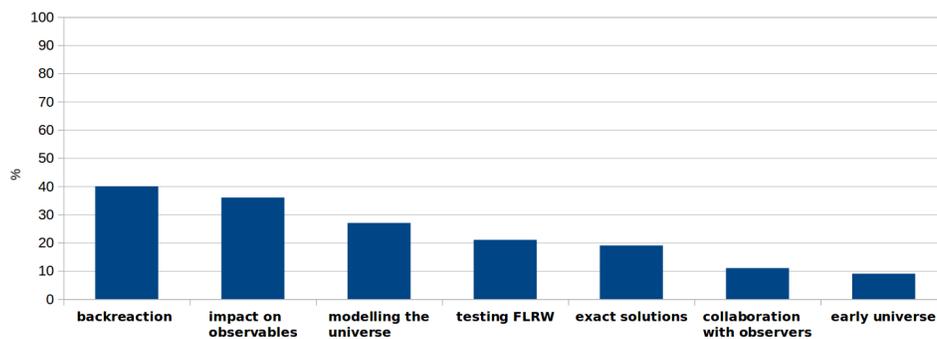}
\caption{{\em Within the domain of inhomogeneous cosmology which topic (or topics) are in your opinion the most important?}}
\label{q2-fig}
\end{center}
\end{figure}

The second question dealt with the current research within the domain of inhomogeneous cosmology, see Fig. \ref{q2-fig}.
The answers we received are as follows:

\paragraph{Backreaction:}

40\% pointed out that the most important topic of inhomogeneous cosmology is backreaction and understanding how nonlinearities and inhomogeneities affect the evolution of the Universe and its properties\cite{2008GReGr..40..467B,2011CQGra..28p4007B,2011CQGra..28p4001E} (cf. 22\% in {\em Question 1}).
A few pointed out that the backreaction and averaging is not well-defined and there are many different approaches to averaging\cite{Buchert:1999er,2008JCAP...04..026R,2009PhRvD..80j3503C,2011JCAP...07..008G,2011JCAP...01..012M,Korzynski:2014nna,Green:2014aga, Buchert:2015iva,Green:2015bma,Green:2016cwo,2005PhRvL..95o1102C,2007PhRvD..75d3506C} and therefore some kind of unification scheme is needed.

\paragraph{Impact on observables:}

Slightly more than a third of respondents answered that the crucial element of inhomogeneous cosmology is to study light propagation effects in the inhomogeneous Universe and the impact of inhomogeneities on observables\cite{2012JCAP...05..003B,2012MNRAS.419.1937M,2012EPJC...72.2242R,2013JCAP...12..051L,2014JCAP...03..040T,2014PhRvD..89f3543A,2016IJMPD..2530007B}.

\paragraph{Modelling the Universe and structure formation:}

28\% of respondents agreed that the most important topic is to develop a good model of the Universe and structure formations. Within the domain of concordance cosmology this topic is studied using the perturbative approach or Newtonian $N$-body simulations. Most of the effort is put towards understanding baryonic feedback and its role in the structure formation\cite{2014MNRAS.444.1518V,2015MNRAS.451.1247S}. However, inhomogeneous cosmology could provide an insight into how nonlinear evolution and backreaction affects the formation of structures\cite{2009suem.book.....B}.
Such studies are essential if we want to obtain a model capable of describing the Universe at all scales with high precision. This will also enable us to answer such questions as for example how accurate the current precision cosmology is.

\paragraph{Testing the fundamental assumptions of the FLRW cosmology:}

Inhomogeneous cosmology allows testing various assumptions fundamental to the concordance cosmology, in particular the spatial homogeneity and isotropy of the Universe.
The assumption of homogeneity and isotropy seems to work well. Furthermore such results as the Ehlers-Geren-Sachs (EGS) theorem \cite{1968JMP.....9.1344E} and the `almost EGS theorem' \cite{1995ApJ...443....1S}, which
make use of the isotropy of the CMB, provide further support for the large scale spatial homogeneity of the Universe. Also observations of the Sunyaev-Zel'dovich effect seem to support large-scale homogeneity \cite{2012PhRvL.109e1303C} and rule out the existence of Gpc--scale inhomogeneities.\cite{2010PhRvD..82j3532F,2011PhRvD..83j3515M,2012PhRvD..85b4002B,2012JCAP...10..009Z,2014A&A...570A..63R}  Still these theories may not be applicable to the real Universe \cite{2009PhRvD..79l3522R}. In addition there are other features of the FLRW models besides the homogeneity and isotropy that still need to be tested. These are for example  fixed and non-evolving curvature ($\partial_t \, k= 0 $, where $k$ is the curvature index and is proportional to $\Omega_k$)\cite{2008PhRvL.101a1301C,2009PhRvD..79h3011L,2009PhRvD..80l3512W,2015PhRvL.115j1301R} or  uniform expansion rate ($\partial_r \, H_0 = 0$)\cite{Romano:2014iea,2014PhRvL.112v1301B,2015arXiv151207364B,2016arXiv160904081E}.
 Therefore, slightly more than a fifth of respondents agreed that one of the most important tasks of inhomogeneous cosmology is to test the assumptions of the FLRW models and to develop predictions of observational signatures that deviate from the FLRW's behaviour.

\paragraph{Studying the exact inhomogeneous solutions of the Einstein equations:}

Most research that uses exact solutions of the Einstein equations is still based on the simplest models such as the Lema\^itre-Tolman models. Even within the LT models the ones that are studied are among the simplest ones with a homogeneous bang time function.
Apart from the LT models, other inhomogeneous models that have been used in cosmological research include the Szekeres models\cite{2006PhRvD..73l3508B,2007PhRvD..75d3508B,2009GReGr..41.1585B,2010PhRvD..82j3510B,2011PhLB..697..265B,2011PhRvD..83h3503K,2011JCAP...05..028N,2012MNRAS.419.1937M,2013PhRvD..87b3524B,2013JCAP...12..051L,2014PhRvD..90l3536P,2014JCAP...03..040T,2015PhRvD..92h3533S,2016JCAP...03..012S}, models with plane symmetry\cite{2014PhRvD..89f3543A}, Lema\^itre models\cite{2006MNRAS.370..924B,2008MNRAS.391L..59B,2010GReGr..42.1935A,2010CQGra..27c5011L}, Stephani models\cite{2007PhRvD..75l3524D,2015PhRvD..91h3506B}, and multi-fluid models\cite{2012JCAP...01..025M}.
It does not help that there is only a handful of exact inhomogeneous solutions that are applicable to cosmology\cite{1997icm..book.....K,Bolejko:2011jc}. Therefore, more effort should be put towards exploring the known solutions and testing the backreaction within these solutions\cite{PhysRevD.19.1058,2004GReGr..36..387P,2008JCAP...10..003B,2012PhRvD..85j3512B,PhysRevD.89.044033,PhysRevD.94.024059}. Another important issue is to analyse cosmological observations within these models\cite{2013A&A...558A..15I,2014A&A...563A..20I}, 
as well as trying to develop and find new solutions with lesser symmetries\cite{1997icm..book.....K}.

\paragraph{Collaboration with the observers:}

Observational data are not only analysed within the framework of FLRW model, but often assumptions and equations of the FLRW models are used to process the raw data and derive the ``observables''. The community of inhomogeneous cosmology should therefore develop a dialogue with the observational community, firstly -- in order to make them aware of the alternatives and help to analyse the data in a less model-dependent way; and secondly -- many relativists are not aware of the wealth of observational data and are not familiar with the observational constraints coming from the astronomical data.
Both communities would therefore benefit from such a dialogue, and 10\% of respondents pointed out that such a dialogue and collaboration is needed and important.

\paragraph{Early Universe:}

Almost 10\% of answers mentioned that the studies of the early Universe (cf. 38\% in {\em Question 1}) should fall within the domain of inhomogeneous cosmology. Topics explicitly mentioned included: alternative solutions to the horizon problem and how inhomogeneities affect the onset of inflation\cite{2013PhRvD..88f3529B}; treating backreaction in the primordial plasma and evaluating the conditions at the last scattering in models where the evolution is not Friedmannian\cite{2015PhRvD..91f3519N}; studies of the spikes and the BKL--type evolution\cite{2005PhRvL..94e1101A,2008CQGra..25d5014L,2012PhRvL.108s1101C,2014CQGra..31a5020L}.

\subsection{Question 3:  How do you think the field of inhomogeneous cosmology will (or should) develop over the next 5--10 years?}

\begin{figure}[h]
\begin{center}
\includegraphics[scale=0.245]{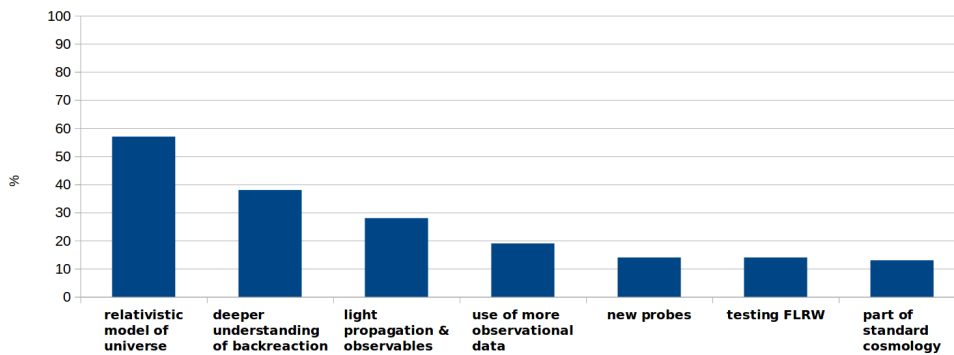}
\caption{{\em How do you think the field of inhomogeneous cosmology will (or should) develop over the next 5--10 years?}}
\label{q3-fig}
\end{center}
\end{figure}

The third question dealt with future avenues of inhomogeneous cosmology and backreaction, see Fig. \ref{q3-fig}. We find the answers and opinions to this question highly valuable as they provide special insight into our field and prospects for its future development. The answers we received are as follows:

\paragraph{Developing a relativistic model of the Universe and structure formation, as well as a deeper understanding of backreaction: }
In {\em Question 2} most people argued that it is backreaction that is currently the main topic of research. It was also noted that the results obtained so far have not been very conclusive. There are many different approaches and results discussed in the literature. As a result it is still unknown how big exactly the effects of backreaction are.
Therefore most respondents pointed out that it will be the numerical relativity that will provide a definite answer to the amplitude and the overall effect of backreaction. The numerical simulations could be developed in several ways:
(i) studying full numerical relativity, and solving the Einstein equations for a cosmological set-up\cite{Bentivegna:2015flc,Mertens:2015ttp,2016arXiv161105447M}; (ii)  developing higher order perturbation theory, especially the relativistic
Lagrangian perturbation theory\cite{2012PhRvD..86b3520B,2013PhRvD..87l3503B,2015PhRvD..92b3512A,2016JCAP...01..030V}; (iii) applying post-Newtonian corrections to $N$-body simulations\cite{2013PhRvD..88j3527A,2014CQGra..31w4006A,2016JCAP...07..053A}.

By developing these methods and performing numerical simulations we will gain a better insight into the backreaction effects.

\paragraph{Light propagation and the effect on observables:}
A large percentage of people argued that in the next few years most efforts should be put towards studying the effects on inhomogeneities on light propagation and observables\cite{2009PhRvD..80l3020K,2010PhRvL.105l1302A,2012JCAP...05..003B,2012EPJC...72.2242R,2012JCAP...04..036B,2013PhRvD..87l3526F,2013JCAP...06..002B,2013PhRvL.110b1301B,2014JCAP...03..040T,2014PhRvD..90l3536P}. Even if the effects turned out to be small and at the percent level, still these studies would be vital for the analysis of future observations. One respondent even provided an analogy with particle physics, where each detector used in the experiment has well-tested characteristics and response; similarly inhomogeneous community could develop catalogue of known astronomical structures and their effect on light passing by.

\paragraph{Better use of the wealth of available observational data:}

Almost 20\% expressed the opinion that the development of inhomogeneous cosmology in the coming years could greatly benefit from inclusion of the constraints coming from the observational data (see also answer to {\em Question 2}).

\paragraph{Developing new observational probes based on the insight from inhomogeneous cosmology and continuing
to test the fundamental assumptions of the FLRW cosmology:}

A small but nonnegligible part of the community argued that we should continue to use inhomogeneous models  in the coming years to develop new relations for observational quantities and predict effects that would otherwise be not observable in the FLRW models either for the sake of learning more about our Universe or for the purpose of testing the fundamental assumptions of the standard cosmological models.
Such tests could include for example testing the large scale homogeneity with the astronomical data\cite{2014MNRAS.443..241W}, testing the effects of  backreaction \cite{2009PhRvD..79h3011L,2009PhRvD..80l3512W,2010PhRvD..81f3515K}, the evolution of spatial curvature\cite{2008PhRvL.101a1301C,2009PhRvD..79h3011L,2009PhRvD..80l3512W,2015PhRvL.115j1301R}, differential expansion of the Universe\cite{Romano:2014iea,2015arXiv151207364B}, and its redshift dependence\cite{2010PhRvD..81h3537S}, or the environmental dependence of the BAO peak\cite{2015MNRAS.448.1660R,2016MNRAS.456L..45R},
the anisotropy pattern of cosmic flow\cite{2012PhRvD..86f3528M,2015arXiv151207364B}, or  decoupling of the geometrical spatial curvature from the dynamical spatial curvature\cite{2012PhRvD..85d3506C}.

\paragraph{Becoming a  part of the concordance cosmology:}

Slightly more than 10\% of respondents expressed an opinion
that inhomogeneous cosmology will in the coming years become part of the standard cosmology with respect to dealing with the effects of inhomogeneities.
If inhomogeneous cosmology is to become a tool
of standard cosmology, we will need to engage in collaborations with observers. Most observers are open-minded
but often not aware of various predictions of relativistic cosmology other than those derived from the FLRW models.
Thus, it is in the interest of inhomogeneous cosmology to collaborate with observational astronomers rather than to separate from them.
This could be achieved by providing either observational predictions or better by providing
free-licensed, clearly written, well-documented software
with relativistic calculations for mainstream cosmologists, which could then be used to analyse the data.

\section{Conclusions}

In 1970 Alan Sandage wrote that cosmology is about two numbers ($\Omega$ and $H_0$) \cite{1970PhT....23b..34S}.
Twenty years later galaxy redshift surveys and the observations of CMB revealed that 2 numbers are not sufficient, and that the minimal number of parameters needed to fit the observations is 6, i.e. 4 parameters to describe the homogeneous background and 2 to describe scalar perturbations. Ten years later we now include at least 2 or 3 more parameters, such as neutrinos, radiation, and tensor perturbations.

More parameters are successively added to the standard cosmological model and some of these describe inhomogeneities (although at the linear level).
In the coming years we are expecting high precision data from DESI, Euclid, LSST, and other observational projects. It is envisaged that  oversimplified models will no longer be applicable to analyse the data at the desirable level of precision and accuracy.

Inhomogeneous cosmology aims at testing the effects of the structure formation and nonlinear evolution on the properties of our Universe.
It is anticipated that in the near future we will be able to
definitively answer the question of whether backreaction effects are significant or
instead are small but nevertheless necessary to analyse the data with high accuracy (such as the recent inclusion of neutrinos is important to analyse the CMB data at percent level precision).

However, without the proper methods, relativistic models, and a dialogue with observers this may not come any time soon. The conclusion therefore is to continue the work and develop the dialogue with the observational community in order to test our models and predictions with the wealth of good quality observational data, and to be sure that no hidden assumptions of the FLRW models go into the processing of the data.

\section*{Acknowledgments}

We are extremely gratefully and would like to thank the 50 academics who kindly agreed to answer our survey.
We also thank Thomas Buchert, Alan Coley, David Wiltshire and Boud Roukema for their comments and suggestions. KB acknowledges the support from the Australian Research Council through the Future Fellowship FT140101270.

\bibliographystyle{ws-procs975x65_eprint}
\bibliography{kbmk-de2}

\end{document}